\documentclass[reprint,superscriptaddress,amsmath,amssymb,aps,floatfix,prl]{revtex4-2}
\usepackage{amsmath}
\usepackage{graphicx} 
\usepackage{dcolumn}  
\usepackage{bm}       
\usepackage{xcolor}   
\usepackage{hyperref} 
\usepackage{upgreek}
\usepackage{comment}

\begin{document}

\title{An optically defined phononic crystal defect}
\author{Thomas J. Clark}
\email{tommy.clark@mail.mcgill.ca}
\author{Simon Bernard}
\author{Jiaxing Ma}
\affiliation{Department of Physics, McGill University, Montréal, Québec H3A 2T8, Canada}
\author{Vincent Dumont}
\affiliation{Laboratory for Solid State Physics, ETH Zurich, CH-8093 Zurich, Switzerland}
\author{Jack C. Sankey}
\email{jack.sankey@mcgill.ca}
\affiliation{Department of Physics, McGill University, Montréal, Québec H3A 2T8, Canada}

\date{\today}

\newcommand{\Om}{$\Omega_{\mathrm{m}}$}
\newcommand{\Gm}{$\Gamma_{\mathrm{m}}$}
\newcommand{\tm}{$\tau_{\mathrm{m}}$}
\newcommand{\um}{$\upmu$m}
\newcommand{\uW}{$\upmu$W}
\newcommand{\SiN}{Si$_3$N$_4$}
\newcommand{\rtHz}{$\sqrt{\text{Hz}}$}

\newcommand{\lr}[1]{\langle #1 \rangle}
\newcommand{\?}[1]{\textcolor{red}{[[~#1~]]}}
\newcommand{\nb}[1]{\textcolor{blue}{[[~#1~]]}}

\begin{abstract}

We demonstrate a mechanical crystal with an optically programmable defect mode. By applying an optical spring to a single unit cell of a phononic crystal membrane, we smoothly transfer a single mechanical mode into the bandgap, thereby localizing its spatial profile from one spanning the entire crystal to one confined within a few unit cells. This localization is evidenced by an enhanced mechanical frequency shift commensurate with a 37-fold reduction in the mode's participating mass. Our results lay groundwork for a new class of optomechanical systems that control mechanical mode profile and participating mass.
\end{abstract}
\maketitle

%
%
%
%




\textit{Introduction.---}
Disrupting periodicity in crystalline systems generates localized ``defect'' modes, which have historically provided indispensable functionalities in condensed matter, optical, electronic, and acoustic systems.  
Within the field of optomechanics, phononic crystal defects are used to isolate a mechanical mode from lossy boundaries \cite{eichenfield2009optomechanical,yu2014a} and ``soft clamp'' \cite{tsaturyan2017ultracoherent, reetz2019analysis} to reduce bending losses. The resulting defect modes exhibit losses low enough to observe multimode squeezing \cite{nielsen2017multimode}, exert measurement-based quantum control \cite{rossi2018measurement}, achieve sensitivities beyond the standard quantum limit \cite{mason2019continuous}, attain ultralong coherence times \cite{kristensen2024long} in the ground state \cite{seis2022ground}, and even observe quantum effects at room temperature \cite{huang2024room}. 
%
Furthermore, phononic crystals -- and the defects they host -- can be engineered. For example, multiple defects can be produced in the same crystal 
\cite{catalini2020soft} and their intermode coupling can be tuned for sensing and transduction applications \cite{catalini2020soft,halg2021membrane}. 
Alternatively, a co-localized \textit{photonic} crystal defect can be introduced \cite{eichenfield2009optomechanical} (potentially in 2D \cite{gavartin2011optomechanical, safavi2010optomechanics, safavi2014two}) to enahnce the optomechanical coupling rate. This enhancement facilitates laser cooling to the quantum ground state \cite{chan2011laser}, observation of quantum motion \cite{safavi2012observation}, and manipulation of individual photons and phonons \cite{cohen2015phonon, meenehan2015pulsed, hong2017hanbury, riedinger2016non}. 
%
%
It is also possible to employ arrays of such defects to realize optomechanically slowed light \cite{chang2011slowing}, synchronization physics \cite{heinrich2011collective}, reservoir engineering \cite{tomadin2012reservoir}, Dirac physics \cite{schmidt2015optomechanical}, Anderson localization \cite{roque2017anderson}, many-body phenomena \cite{ludwig2013quantum} including topological effects \cite{peano2015topological, wang2015topological, brendel2018snowflake}, modified \cite{chen2014photon} or non-reciprocal \cite{seif2018thermal} transport, and the generation of synthetic fields for photons \cite{schmidt2015optomechanical, walter2016classical, zapletal2019dynamically} and phonons \cite{brendel2017pseudomagnetic}.

The current paradigm for creating mechanical crystal defects relies on careful simulation and precise lithography to create a geometric disturbance in the crystal structure. More importantly, once fabricated, the mass and mode profile of such defects is fixed by the geometry that was physically realized, requiring multiple iterations to find a device having the desired properties. Here we instead fabricate a simple uniform phononic crystal and then \emph{optically} disrupt the periodicity to create a solitary, programmable defect mode with widely tunable properties. This effect is fully reversible, and the resulting defect has a smoothly tunable degree of localization that can in principle be adjusted as fast as the optical power can be modulated. Our proof-of-concept demonstration naturally extends to programmable dimers, waveguides, lattices, and more, greatly expanding the range of possible optomechanical systems and studies.

\begin{figure}[ht]
\centering
  \includegraphics[width=0.5\textwidth]{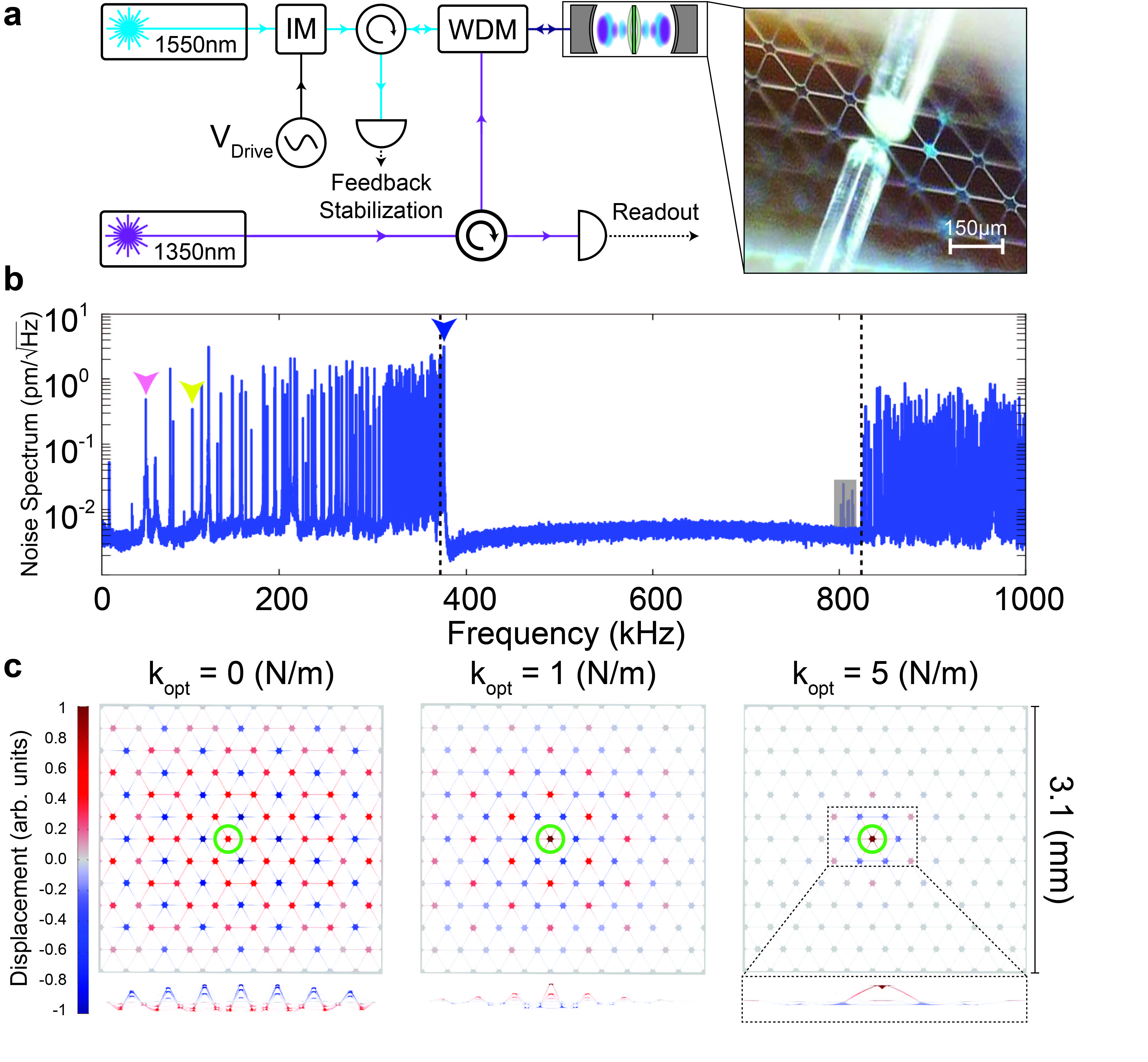}
  \caption{ \label{fig1} Apparatus for observing an optically controlled mechanical crystal defect.
  (a) A 1550 nm ``control'' beam passes through an intensity modulator (IM), circulator, and wavelength division multiplexer (WDM) into a fiber cavity centered on a phononic crystal lattice site (inset image). Radiation forces can be applied via the IM control voltage $V_\text{drive}$, while a second 1350 nm (purple) beam provides independent, backaction-free readout of membrane motion. The reflected 1550 nm light is used to stabilize the laser's detuning from the cavity resonance via feedback to shear piezos attached to the fiber mirrors. 
  (b) Thermally driven mechanical noise spectrum. Vertical dashed lines indicate the limits of the band gap simulated by a finite element solver (COMSOL). Markers indicate the individual modes at the focus of this study, with pink being the fundamental (1,1) membrane mode (54 kHz, simulated mode mass 149 ng), gold is the (3,1) mode (116 kHz, 173 ng), and blue is the band edge mode (375 kHz, 162 ng). The gray box highlights a region where ``edge-like modes'' are predicted to (faintly) appear.
  (c) Membrane geometry and simulated displacement profile of the band edge mode with 0 N/m, 1 N/m, and 5 N/m optical spring applied to a central lattice site (circled). The $\sim$5 $\upmu$m $1/e^2$ radius perturbation breaks the crystal's periodicity, producing a localized (bound) defect mode with reduced participating mass.}
\end{figure}

\textit{Optomechanical Platform.---}
The apparatus used to generate and measure an optically defined mechanical defect is shown in Fig.~\ref{fig1}. 
A uniform phononic crystal membrane (3.3~mm $\times$ 3.1~mm, 180 nm thick Si$_3$N$_4$ patterned into a hexagonal lattice) is positioned within a 30-$\upmu$m-long, finesse $\sim10^4$ fiber optical cavity \cite{hunger2010fiber} such that the cavity mode intersects the center of a single unit cell. Light in the cavity can then apply a point-like optical spring normal to the membrane -- this can be formally viewed as a potential well in the membrane's wave equation \cite{2016barasheed-optically} -- thereby disrupting the crystal's periodicity and creating a mechanical bound state.  
The optical fibers are epoxied \cite{saavedra2021tunable} within tight-tolerance ferrules fixed to shear piezos, enabling independent cavity length and position adjustment. To reduce mechanical dissipation and acoustic noise, the entire apparatus is housed within a vibration-isolated ultra-high vacuum chamber at pressure $2 \times 10^{-8}$ Torr. 
Figure~\ref{fig1}(a) shows the optical circuit, which employs a 1550 nm ``control'' beam (blue) to exert the optical spring. 
This beam can also measure membrane motion, and Fig.~\ref{fig1}(b) shows the thermally driven spectrum (at low power), which exhibits a well-defined bandgap with boundary frequencies near those predicted by a finite-element simulation (dashed lines). For calibration, we employ a second ``readout" beam (purple in Fig.~\ref{fig1}(a)) with a wavelength (1350 nm) outside the high-reflectivity band of the mirrors; this provides a low-finesse readout with negligible backaction, and (most importantly) a sensitivity that is independent of control beam dynamics and optomechanical noise interference \cite{weis2010optomechanically, nielsen2017multimode}. 
Figure~\ref{fig1}(c) shows the expected evolution of a band edge mode (blue marker in (b)) from fully delocalized (left) to localized (right), as the  power (applied to the circled ``pad'') is increased.

\begin{figure}[ht] 
    \centering
   \includegraphics[width=0.48\textwidth]{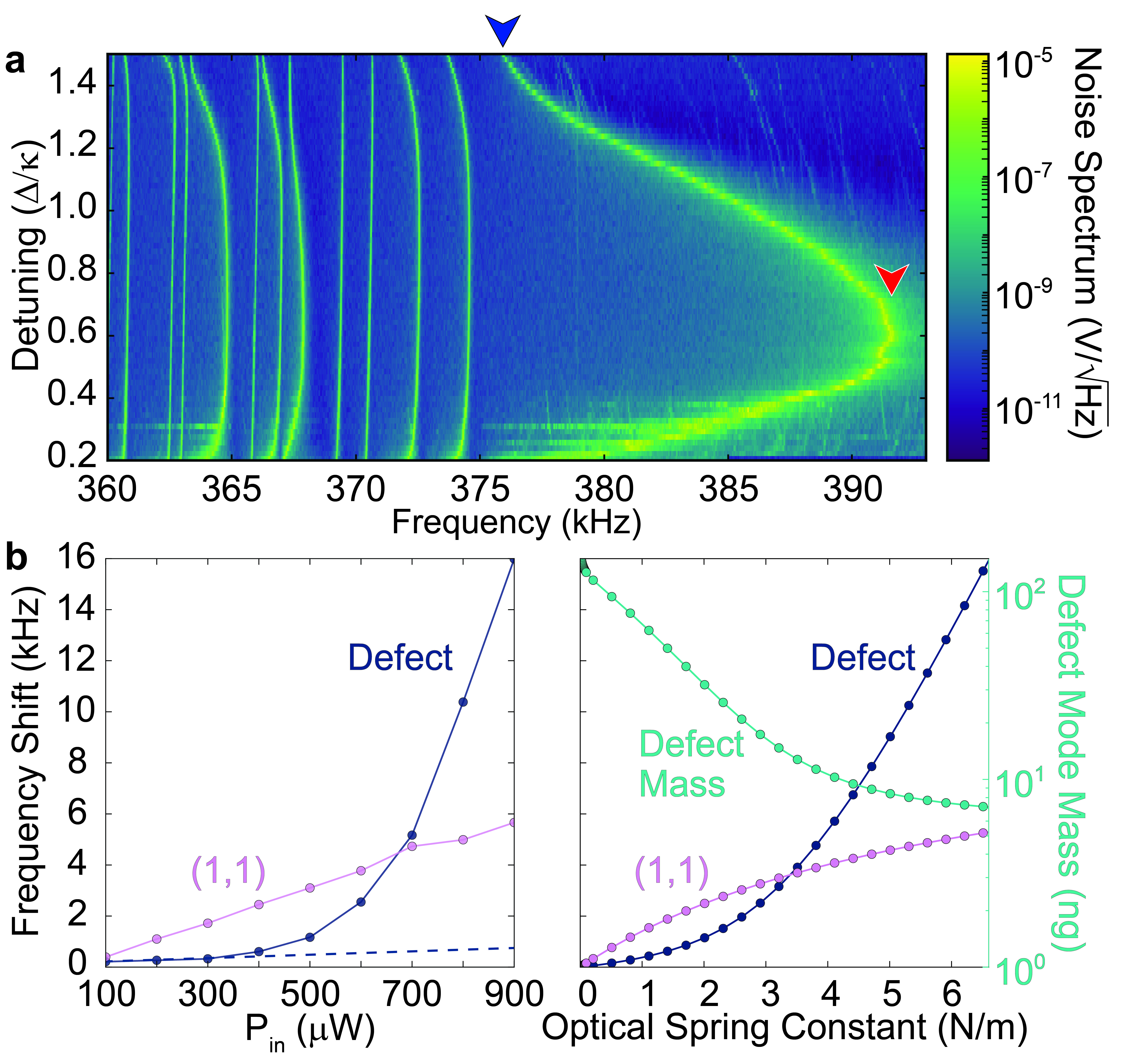}
\caption{ \label{fig2} Evidence of defect localization in the frequency domain.
   (a) Dependence of thermally driven mechanical spectrum on control beam (incident power $\sim$ 0.84 mW) detuning $\Delta$ (normalized by the cavity decay rate $\kappa$) near the lower edge of the band gap. The frequency shift of the band edge mode greatly exceeds that of the other modes, consistent with the hybridization predicted for many band edge modes optically stiffened and coupled to one another.
   (b) Measurement (left) and simulation (right) of frequency shifts for the band edge (blue) and fundamental (1,1) (pink) modes (following the colors in Fig.~\ref{fig1}(b)). The frequency shift of the fundamental mode is in stark contrast to with the superlinear response of the band edge mode, which shifts 24.1 times farther than what is predicted for a single mode of fixed mass; the dashed blue line extrapolates the initial linear dependence for reference. The teal curve shows the simulated effective mass dropping to $\sim$ 4 \% of the untrapped value. Error bars (standard error of the mean) are smaller than the data points.}
\end{figure}

\textit{Enhanced Frequency Shift.---}
One hallmark of optically induced localization is an anomalously large frequency shift in response to an applied optical spring \cite{2016barasheed-optically}. This is demonstrated in Fig.~\ref{fig2}(a), which shows how the mechanical spectrum near the lower band edge varies with control beam's detuning from cavity resonance $\Delta$. At the detuning of maximal spring (red marker), the frequency of a single bright mode has shifted $\sim$20 times farther than the other modes (into the gap). 
Approaching this optimal detuning, the other modes exhibit a complex network of avoided crossings, consistent with optically mediated hybridization; this is not surprising, since each photon not only applies a spring to each mode, but also simultaneously couples every mode to every other mode with a comparably stiff spring \cite{supplementary}. In fact, for $N$ participating modes, there are $\sim N^2$ inter-mode springs \cite{supplementary} that collectively generate the single mode with enhanced response observed in Fig.~\ref{fig2}(a). 
If the modes were not so densely spaced or they otherwise did not hybridize, we would expect all of them to shift according to their nominal material spring $K_\text{m}$ mode mass $m$, and optical spring $K_\text{opt}$ as
\begin{equation}\label{eq:frequency-shift-naive}
\Omega = \sqrt{\frac{K_\text{m}+K_\text{opt}}{m}}\approx \Omega_\text{m} + \frac{K_\text{opt}}{2 m\Omega_\text{m}}
\end{equation}
(where $\Omega_\text{m}=\sqrt{K_\text{m}/m}$ is the nominal frequency), i.e., roughly linearly with $K_\text{opt}$. To illustrate this difference more quantitatively, Fig.~\ref{fig2}(b) compares how the band edge mode (blue) and fundamental (1,1) mode (pink) frequencies shift with increasing incident power $P_\text{in}$. The comparatively isolated fundamental mode resonance frequency (nominally 54 kHz) shifts roughly in proportion to the optical spring constant ($\propto P_\text{in}$), as expected for a single harmonic oscillator of mass 149 ng (simulated). 
Meanwhile, the band edge mode (nominally 375 kHz, 162~ng) is found to initially shift a factor of 0.084 as far, consistent with the expected factor of $\sim$ 0.13 from $m$ and $\Omega_\text{m}$ in Eq.~\ref{eq:frequency-shift-naive} and an optomechanical coupling that is 0.8 times that of the fundamental (due, e.g., to the mode's different profile within the cavity light). 

However, with increasing power, the band edge mode rapidly pulls away from its extrapolated linear dependence (dashed line), producing a final frequency shift that is, despite the above suppressing factors, 2.8 times larger than that of the fundamental and 24 times larger than the extrapolated value. This enhanced response signals the collective interaction of the densely populated modes near the band edge. The confirming simulation (right) predicts a mode mass mass (teal) reduced to 4\% of the unperturbed, delocalized value.
 
\begin{figure}[ht]
    \centering
   \includegraphics[width=0.48\textwidth]{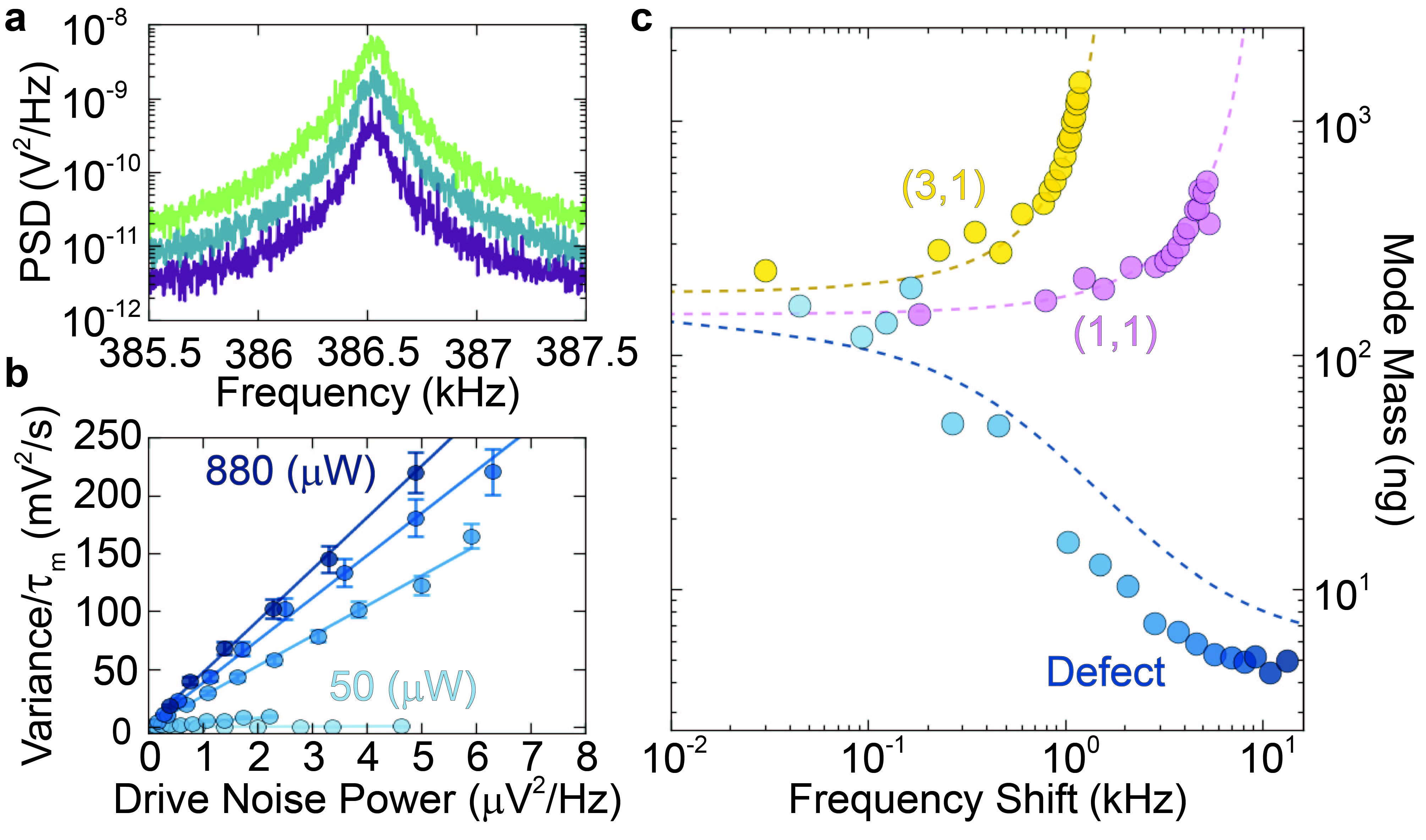}
\caption{\label{fig3} 
   Force-driven measurement of mechanical mode mass.
   (a) Noise spectra of the band edge mode driven by $P_\text{in}=880$ $\upmu$W~at the optimal detuning, with optical white noise drive having (measured) amplitude spectral densities 3.5 $\upmu$W/$\sqrt{\text{Hz}}$~ (purple, lowest), 6.8 $\upmu$W/$\sqrt{\text{Hz}}$~ (teal), and 12.7 $\upmu$W/$\sqrt{\text{Hz}}$~ (green, highest), using the fixed-sensitivity 1350 nm beam for readout.  
   (b) Linear scaling of integrated defect mode resonance noise with input noise power (i.e., the variance) scaled by the mechanical ringdown time $\tau$ for incident powers $P_\text{in}=$ 50 (faintest, bottom), 350, 600, 750, 880  (darkest, top) $\upmu$W. Plotted this way, linear fits (lines) with larger slope correspond to modes with lower mass \cite{supplementary}. Error bars represent standard error of the mean. 
   (c) Mass of the three modes indicated in Fig.~\ref{fig1}(b), -- estimated from the fit slopes in (b) -- as a function of optical spring strength. The strength is expressed as a frequency shift of the stiffened modes. Unperturbed values are estimated from the simulation at zero power (dotted lines), which sets a constant overall scale for each mode. The (1,1) and (3,1) modes (pink and gold) exhibit increasing mass due to ``pinning'', while the defect mode mass is strongly reduced (despite pinning) due to localization.
   }
\end{figure}

\textit{Defect Mode Mass.---}
We can also more directly measure the evolution of the mode mass $m$ by driving its motion with radiation forces and measuring its response, exploiting the fact that a harmonic oscillator (with or without an optical spring) has susceptibility
\begin{equation}
    \chi(\omega) = \frac{1/m}{\Omega^2-\omega^2 + i2\omega/\tau} ,
\end{equation}
scaling inversely with $m$, where ($\Omega$, $\tau$) are the separately measured resonance frequency and amplitude ringdown time. To avoid artifacts from frequency drifts, we apply a known force noise power spectral density (PSD)  $S_F$ near $\Omega$ via the intensity modulator \cite{kumar2023novel}, and measure the resulting displacement variance with the readout beam. Figure~\ref{fig3}(a) shows readout noise spectra near the defect mode frequency with control beam power $P_\text{in}=880~\upmu$W and an applied optical noise $\sqrt{S_P}$ of 3.5~$\upmu$W/$\sqrt{\rm{Hz}}$ (bottom), 6.8~$\upmu$W/$\sqrt{\rm{Hz}}$ (middle), and 12.7~$\upmu$W/$\sqrt{\rm{Hz}}$ (top); these noise values are directly measured with a pick-off diode after the intensity modulator, and notably include the laser's inherent classical noise. Each spectrum exhibits a resonant peak following the expected proportionality to $|\chi|^2S_P~  (\propto |\chi|^2S_F)$, such that the integrated spectrum (i.e., the variance) $\propto \tau S_P/m^2$ \cite{supplementary}, providing a simple estimator for changes in $m$. Figure~\ref{fig3}(b) shows this variance, scaled by $\tau$ to emphasize changes in $m$, for swept $S_P$ and multiple values of $P_\text{in}$ from 50~$\upmu$W~(bottom, lightest data) to 880~$\upmu$W~(top, darkest data). When plotted this way, decreased mode mass $m$ presents as an increased slope. Figure~\ref{fig3}(c) shows the evolution of the so-measured mode mass for more values of $P_\text{in}$. The statistical uncertainty of this measurement is too small to be seen, and the systematic fluctuations at low power reflect real changes in mode mass, likely arising from hybridization with the modes of the supporting structure \cite{ni2012enhancement} (the silicon chip), which is not included in the simulation; consistent with this intuition, these fluctuations are largest when the mode is more delocalized and can interact more strongly with the supporting frame. This unpredictable behavior -- and our ability to tune around it -- highlights a major advantage of this approach. 

Also included is the same measurement for two ``reference'' modes, the fundamental (1,1) (pink) and (3,1) (gold) membrane modes indicated in Fig.~\ref{fig1}(b), along with simulated values (dashed lines) for each. The agreement between the longer-wavelength reference modes and simulation further validates the mass estimation method. The factor of $\sim$2 discrepancy between the simulated and actually measured defect mass (a mode having significantly smaller wavelength) likely reflects our lack of knowledge about small-scale fabrication imperfections not included in the simulation. 

The most important feature of Fig.~\ref{fig3}(c) is the observed trends: the reference modes both exhibit an \emph{increasing} effective mass, as expected for a point-like optical spring that ``pins'' the membrane amplitude \cite{ni2012enhancement,chang2012ultrahigh,muller2015enhanced,2016barasheed-optically}, and (despite this pinning effect) the measured bound state mode mass drops to 2.7\% of its nominal value. This observation of an optically localized mechanical defect mode is the main result of this work.

\begin{figure}[h]
  \centering
\includegraphics[width=0.4\textwidth]{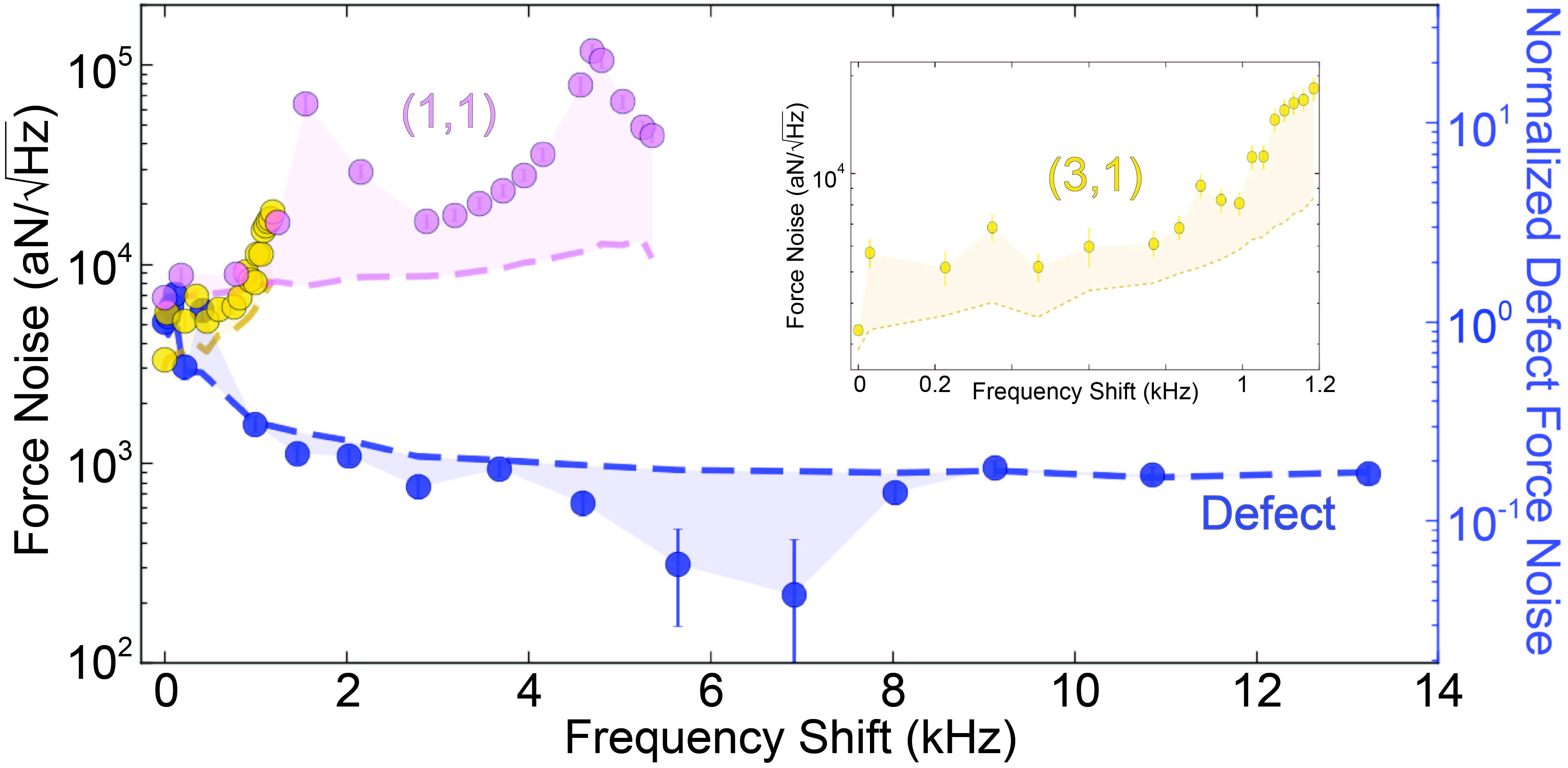}
\caption{ \label{fig4}
    Measured force noise at zero applied drive, extracted from the fit offsets in the variance (i.e., Fig.~\ref{fig3}(b)). Dotted lines indicate the expected thermal force noise for the observed change in mass and spring, for fixed intrinsic damping. Error bars represent propagated uncertainty from the fits in Fig.~\ref{fig3}(b). The defect mode (blue) exhibits force noise significantly below the constant damping expectation at certain optical spring strengths, highlighting certain mode shapes and frequencies that are better isolated from the thermal environment.  By contrast, the (1,1) and (3,1) (pink and gold) exhibit force noise in excess of the thermal expectation, hinting at an increase in intrinsic damping associated with their changing mode shapes and frequencies. The inset expands the data for the (3,1) mode.  
     }
\end{figure}

\textit{Outlook.---}
Although the chosen geometry of this study is not optimized for sensing, it is still instructive to measure the thermal force noise of the modes as their profiles change shape. Figure~\ref{fig4} shows the force noise spectral density of the three studied modes, estimated by extrapolating the variances (e.g., those in Fig.~\ref{fig3}(b)) to zero noise power, such that only the response to environmental noise remains \cite{supplementary}. Also shown (dashed lines) is the evolution of the thermal noise naively expected from the observed changes in mass and spring alone, assuming a fixed temperature bath and fixed intrinsic mechanical damping \cite{supplementary}. Interestingly, the defect mode's measured force noise is generally lower (and occasionally much lower) than this expectation, likely due to a combination of reduced clamping losses and a varying distribution of bending losses as the mode shape is tuned. By contrast, the reference modes show the opposite trend, notably opposing what one might expect from other partially levitated systems \cite{ni2012enhancement,chang2012ultrahigh,muller2015enhanced}. For these reference modes, pinning plays a dominant role, increasing the perceived mode mass and introducing additional bending losses. The observed peaks suggest additional loss mechanisms to explore, such as coupling to modes of the supporting structure, increasing energy loss when nearly degenerate. This preliminary measurement highlights the technique's potential for untangling dissipation mechanisms in mechanical sensors, motivating nontraditional studies: rather than fabricating many ``identical'' devices, one can now smoothly vary what regions of a single mechanical system are bending during oscillation, and change the degree to which a mode is coupled to lossy boundaries, other modes, and / or deposited materials. 
In a similar direction, one could optically tune \textit{in situ} the overlap (and interaction strength) between modes of defect dimer sensors \cite{catalini2020soft, halg2021membrane, halg2022strong} or with another system of interest, such as a cavity at a different lattice site, a capacitively coupled qubit, or any system that can be force-coupled to a mechanical device, enabling hybrid quantum systems \cite{stannigel2011optomechanical, chu2020perspective,kurizki2015quantum} with optically tuned coupling. Beyond individual defects, with more optical sites (and / or a spatial light modulator), it should furthermore be possible to realize programmable phononic waveguides \cite{hatanaka2014phonon, ren2022topological, Brendel2018Jan}, enabling new architectures for mechanically shuttling signals between different systems within a naturally ultracoherent framework \cite{tsaturyan2017ultracoherent, seis2022ground, kristensen2024long}. Within the context of condensed matter physics, this optical ``potential'' term in the phonon wave equation \cite{2016barasheed-optically} provides a natural testbed for 1D and 2D periodic systems with arbitrary and programmed superlattices, defects, and disorder, while also providing access to qualitatively new studies of Anderson localization \cite{anderson1958absence}, time-varying dispersion/wavefront shaping \cite{hatanaka2019electrostatically}, and topologically protected modes \cite{ren2022topological, lu2017observation}, not to mention explorations of energy transfer (i.e., Landau-Zener dynamics) between bound states and a quasi-continuum \cite{basko2017landau}. We note that, unlike heat-induced loosening of a predefined structural defect \cite{sadeghi2020thermal,shaniv2023direct}, the optical perturbations in our system arise from a coherent radiation pressure interaction that responds on a time scale set by the rate at which the optical power can be modulated; it can thus quickly and reversibly create new structures, and is not limited by steady state heat flow, or related thermal time scales. 
Finally, we recall that, due to the $\sim N^2$ scaling of the optical spring network, larger crystals (which have a higher density of states) do not require stronger optical springs to achieve the same level of localization. This means the per-photon mechanical response can be made arbitrarily large, in principle generating significant response with an average of a single photon in a chip-scale device \cite{2016barasheed-optically}. We suggest that this collective behavior may represent a useful resource on the route to creating exotic macroscopic states of motion. 

\section{Acknowledgements}
We gratefully acknowledge initial fabrication support from Abeer Barasheed. TC acknowledges conversations with Sebastian Spence and Michael Caouette-Mansour, and financial support from the Walter Sumner Fellowship. VD acknowledges financial support from FRQNT-B2 Scholarship and McGill Schulich Graduate Fellowship. JCS acknowledges support from the Natural Sciences and Engineering Research Council of Canada (NSERC RGPIN 2018-05635), Canada Research Chairs (CRC 235060), Canadian foundation for Innovation (CFI 228130, 36423), Institut Transdisciplinaire d'Information Quantique (INTRIQ), and the Centre for the Physics of Materials (CPM) at McGill.

\section{Data Availability} 
The data presented here will be made available on the McGill Dataverse found at \\ https://borealisdata.ca/dataverse/mcgill.

\providecommand{\noopsort}[1]{}\providecommand{\singleletter}[1]{#1}%
%


\end{document}